# Absence of $T_c$-Pinning Phenomenon Under High Pressure in High-Entropy $REO_{0.5}F_{0.5}BiS_2$ Layered Superconductor

Yoshikazu Mizuguchi[1]*, Kazuki Yamane[2,3], Ryo Matsumoto[2], Kazuhisa Hoshi[1], Aichi Yamashita[1], Akira Miura[4], Yuki Nakahira[1,5], Hirokazu Kadobayashi[6], Saori I. Kawaguchi[6], and Yoshihiko Takano[2,3]

[1]*Department of Physics, Tokyo Metropolitan University, Hachioji, 192-0397, Japan*
[2]*International Center for Materials Nanoarchitectonics (MANA), National Institute for Materials Science, Tsukuba, 305-0047, Japan*
[3]*University of Tsukuba, Tsukuba, 305-8577, Japan*
[4]*Faculty of Engineering, Hokkaido University, Sapporo 060-8628, Japan*
[5]*Foundational Quantum Technology Research Directorate, National Institutes for Quantum Science and Technology (QST), Hyogo 679-5148, Japan*
[6]*SPring-8, Japan Synchrotron Radiation Research Institute, Hyogo, 679-5198, Japan*

Recently, robustness of superconductivity (transition temperature, $T_c$) under high pressures has been observed in high-entropy alloy (HEA), bcc-type Ti-Zr-Hf-Nb-Ta, and HEA-type compounds (Ag,In,Sn,Pb,Bi)Te with a NaCl-type structure. Since those materials have three-dimensional crystal structure, investigation on the pressure dependence of $T_c$ of low-dimensional materials is needed to understand the phenomena. Here, we investigated the superconducting properties and the crystal structure of $BiS_2$-based layered system $REO_{0.5}F_{0.5}BiS_2$. Although the robustness of $T_c$ was induced in $M$Te with increasing $M$-site configurational entropy, the increase in RE-site configurational entropy does not induce robustness of $T_c$ under high pressures in $REO_{0.5}F_{0.5}BiS_2$. The crystal structure of HEA-type $REO_{0.5}F_{0.5}BiS_2$ was confirmed as monoclinic $P2_1/m$, which is the same space group as the zero-entropy counterpart $LaO_{0.5}F_{0.5}BiS_2$. The results suggest that an increase in configurational entropy at blocking layers do not affect crystal structure and superconducting properties under high pressures in the $BiS_2$-based layered system.

## 1. Introduction

High-entropy alloys (HEAs) are multi-element alloys containing five or more elements with 5–35% [1,2]. Because of high performance in various mechanical and functional properties, HEAs have been extensively studied in the field of materials science [3]. In 2014, superconductivity in a HEA, Ti-Zr-Hf-Nb-Ta, was reported [4]. After the discovery, many HEA superconductors have been developed [5–12]. One of the noticeable features of HEA superconductors is their superconducting characteristics clearly different from low-entropy crystalline alloys and amorphous alloys [5]. Furthermore, robustness of superconductivity under high pressures was reported in Ti-Zr-Hf-Nb-Ta by Guo et al. where comparable superconducting transition temperatures ($T_c$s) were observed up to about 200 GPa [13]. Later, this $T_c$-*pinning* phenomenon of the HEA was explained by theoretical estimation of electronic density of states [14]. However, in the study, contribution of phonon on superconductivity pairing was not estimated.

Recently, we studied pressure effects on superconductivity, crystal structure, and electronic states of NaCl-type metal telluride $M$Te family ($M$ = Ag, In, Sn, Pb, Bi) with different configurational entropy of mixing ($\Delta S_{mix}$) [15,16]. In zero-entropy PbTe, two-step structural transitions from cubic (NaCl), orthorhombic, to cubic (CsCl) [Fig. 1(a)] is observed [17–19], but the orthorhombic phase was suppressed in medium- and high-entropy-alloy-type $Ag_{1/3}Pb_{1/3}Bi_{1/3}$Te and $Ag_{0.2}In_{0.2}Sn_{0.2}Pb_{0.2}Bi_{0.2}Te_5$ [20]. Therefore, in the simple $M$Te structure, pressure dependence of crystal structure was modified by an increase in $M$-site $\Delta S_{mix}$. The noticeable feature in $M$Te is the observation of $T_c$-pinning states under high pressures at $P$ = 15–35 GPa in a CsCl-type phase of HEA-type (Ag,In,Sn,Pb,Bi)Te, whereas PbTe exhibits a clear decrease in $T_c$ in the CsCl-type phase. [20]. Further, simulations on atomic vibration in $M$Te suggested the emergence of glassy vibrational density of states by an increase in $M$-site $\Delta S_{mix}$ [21]. X-ray absorption spectroscopy [20] and electronic band calculations suggested that electronic density of state in $M$Te with all zero-, medium-, and high-entropy-alloy compositions show similar decrease with increasing pressure. Therefore, the current scenario of the $T_c$-pinning states in $M$Te is the explanation with the glassy phonon and unique electron-phonon coupling in HEA-type $M$Te caused by introduced huge $M$-site disorders [21].

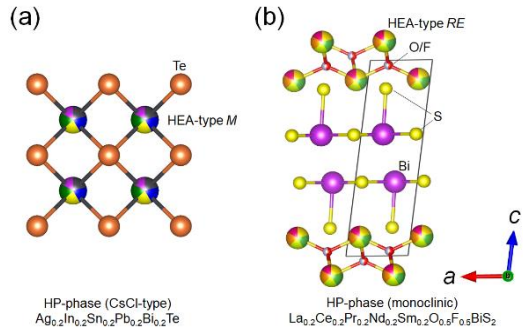

(Color online) Fig. 1. Schematic images of crystal structure of the high-pressure (HP) phase of (a) $M$Te and (b) $RE$O$_{0.5}$F$_{0.5}$BiS$_2$. The solid square indicates the unit cell of the monoclinic structure.

To reach a solid conclusion on the mechanisms of the $T_c$-pinning phenomenon in HE-type materials, further examples with/without $T_c$-pinning sates in the pressure dependence of $T_c$. In this study, we focus on BiS$_2$-based layered superconductors $RE$O$_{0.5}$F$_{0.5}$BiS$_2$ ($RE$ = La, Ce, Pr, Nd, Sm) [Fig. 1(b)] because a previous study revealed that the increase in $RE$-site $\Delta S_{mix}$ results in the suppression of local disorder at the in-plane S site; the positive correlation between the suppression of in-plane disorder and the emergence of bulk superconductivity is an established trend in the system [22–25]. The suppression of structural disorder in the $RE$O$_{0.5}$F$_{0.5}$BiS$_2$ system by an increase in $RE$-site $\Delta S_{mix}$ is a trend contrasting to the case of $M$Te with a variable $M$-site $\Delta S_{mix}$. Therefore, the difference in the evolution of structural disorder by tuning $\Delta S_{mix}$ should depend on crystal structures. Furthermore, in BiS$_2$-based system, the electronic states near the Fermi energy are dominated by Bi-6p electrons in $RE$O$_{0.5}$F$_{0.5}$BiS$_2$ [22], which simply implies that the electronic states of $RE$O$_{0.5}$F$_{0.5}$BiS$_2$ are not affected by an increase in $RE$-site $\Delta S_{mix}$. Therefore, to link the $T_c$-pinning phenomenon and glassy atomic vibrations caused by site disorder, for example in $M$Te, the confirmation of the *absence* of $T_c$-pinning states in HEA-type $RE$O$_{0.5}$F$_{0.5}$BiS$_2$ is important. That should be useful to further investigate the physical properties of HEA-type layered superconductors [26–29]. Here, we show that the increase in $RE$-site $\Delta S_{mix}$ in $RE$O$_{0.5}$F$_{0.5}$BiS$_2$ does not affect the pressure-induced tetragonal-monoclinic structural transition, and the pressure dependences of $T_c$ for all the examined $RE$O$_{0.5}$F$_{0.5}$BiS$_2$ exhibit clear decrease with pressure.

## 2. Experimental Details

Polycrystalline samples of $RE$O$_{0.5}$F$_{0.5}$BiS$_2$ with $RE$ = Pr, Ce$_{0.5}$Nd$_{0.5}$, Ce$_{1/3}$Pr$_{1/3}$Nd$_{1/3}$, La$_{0.2}$Ce$_{0.2}$Pr$_{0.2}$Nd$_{0.2}$Sm$_{0.2}$ were prepared by solid-state reaction as described in Ref. 23. According to the $RE$-site composition, $\Delta S_{mix}$ for the above four compositions is calculated as 0, 0.69$R$, 1.1$R$, and 1.6$R$, respectively.

The high-pressure synchrotron X-ray diffraction (HP-SXRD) was performed at BL10XU of SPring-8 under the proposal (No.: 2022A1086). The powder sample was loaded into a sample hole drilled in a SUS gasket together with a pressure medium, He gas, using 200 MPa gas loading system at SPring-8. The samples were compressed to the pressures of interest using single crystal diamond anvils. The actual pressure generated in the cell was determined by wavelength shift of the ruby R1 fluorescence line [30]. An imaging plate detector Rigaku R-AXIS IV++ was used to collect SXRD patterns where the exposure time was 80 s, and the DAC was oscillated during the experiments. IP Analyzer [31] was used to obtain one-dimensional data from two-dimensional original data. The wave length of X-ray was 0.412868 Å. The obtained HP-SXRD patterns were refined by the Rietveld method using RIETAN-FP [32]. Schematic images of the crystal structures were depicted using VESTA [33].

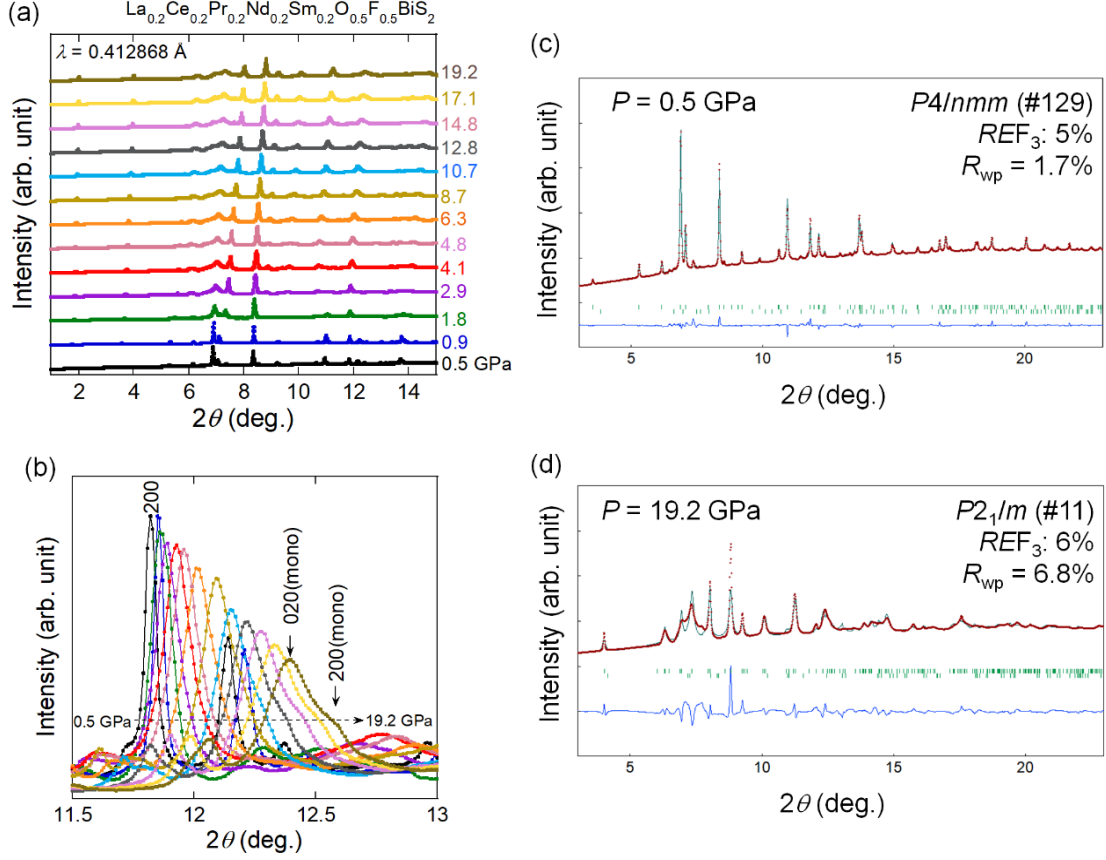

(Color online) Fig. 2. (a) HP-SXRD patterns for HEA-type $La_{0.2}Ce_{0.2}Pr_{0.2}Nd_{0.2}Sm_{0.2}O_{0.5}F_{0.5}BiS_2$. (b) Evolution of the 200 peak (tetragonal) that splits into 020 and 200 peaks in the monoclinic phase. (c,d) Typical Rietveld refinement results for low-$P$ (tetragonal) and high-$P$ (monoclinic) phases.

Electrical resistance measurements were performed at ambient pressure and high pressures using a standard four-probe method on a Physical Property Measurement System (PPMS, Quantum Design). A diamond anvil cell (DAC) with boron-doped diamond electrodes was used for the high-pressure measurements [34,35]. We used cBN for a pressure medium. The applied pressure was estimated from the relationship between the pressure and wavelength of the ruby fluorescence [36] measured by an inVia Raman Microscope (RENISHAW).

## 3. Crystal structure

Figure 2(a) shows the HP-SXDR patterns for $La_{0.2}Ce_{0.2}Pr_{0.2}Nd_{0.2}Sm_{0.2}O_{0.5}F_{0.5}BiS_2$. The HP-SXRD patterns for $P$ = 0.5 and 0.9 GPa could be refined with a tetragonal (*P4/nmm*) moder as shown in Fig. 2(c) with a minor impurity phase of $REF_3$. At $P \geq 1.8$ GPa, clear peak broadenings

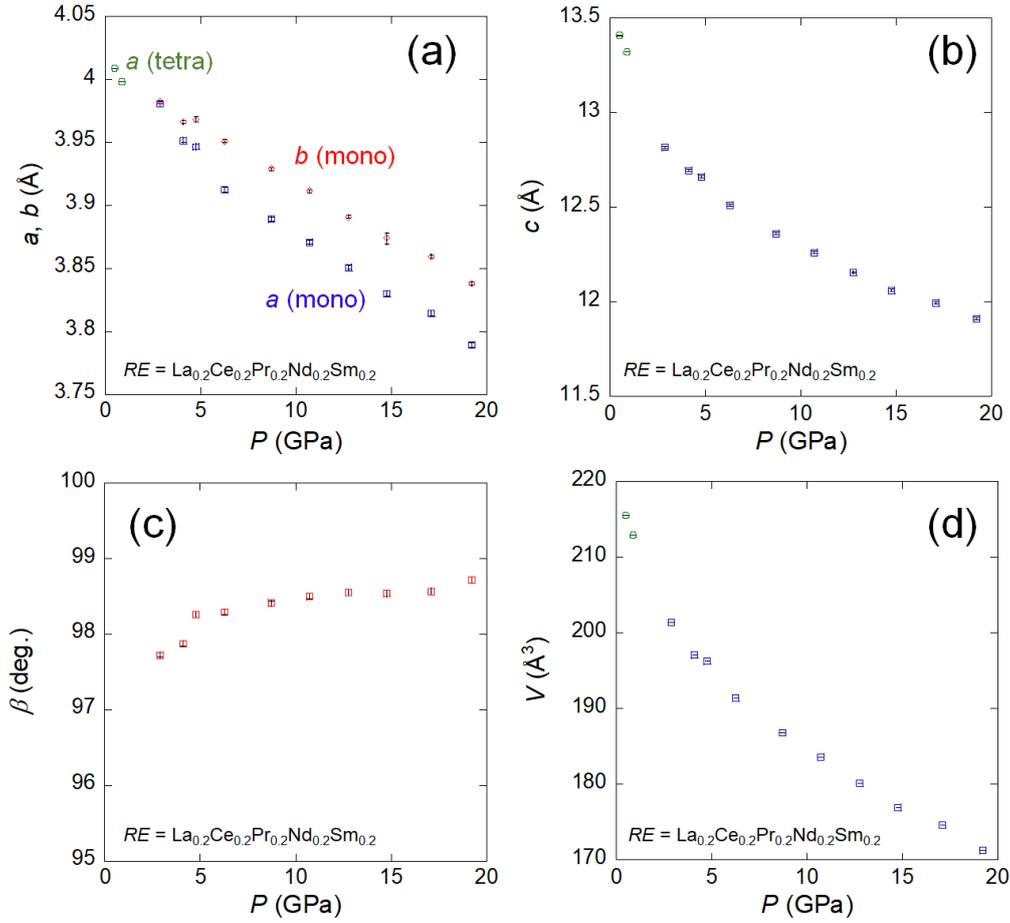

(Color online) Fig. 3. Pressure dependence of (a) lattice parameter $a$ and $b$, (b) lattice parameter $c$, (c) $\beta$ angle, (c), and (d) volume for HEA-type $La_{0.2}Ce_{0.2}Pr_{0.2}Nd_{0.2}Sm_{0.2}O_{0.5}F_{0.5}BiS_2$. In Fig. (a), tetra and mono denote tetragonal and monoclinic, respectively.

were seen, which corresponds to the structural transition to monoclinic ($P2_1/m$). The typical Rietveld refinement result for the high-$P$ phase is shown in Fig. 2(d). Although the fitting is not good due to broadened peaks under high pressure, which has been also seen in a previous work on $LaO_{0.5}F_{0.5}BiS_2$ [37], the peak positions are well explained by the monoclinic model. Therefore, the HEA-type $La_{0.2}Ce_{0.2}Pr_{0.2}Nd_{0.2}Sm_{0.2}O_{0.5}F_{0.5}BiS_2$ also exhibits a pressure-induced structural transition from tetragonal to monoclinic, which is same as zero-entropy counter-part $LaO_{0.5}F_{0.5}BiS_2$ [37]. In $M$Te, the lower-symmetry (orthorhombic) phase was suppressed by an increase in $M$-site $\Delta S_{mix}$, but, in the case of $RE O_{0.5}F_{0.5}BiS_2$, the increase in $RE$-site $\Delta S_{mix}$ less affects the crystal-structure phase diagram under high pressures [37]. The obtained lattice parameters are summarized in Fig. 3. Lattice parameters $a$ and $b$ clearly split after the structural transition. The obtained $\beta$ is close to that reported for $LaO_{0.5}F_{0.5}BiS_2$ under high pressures [37].

## 4. Superconducting properties

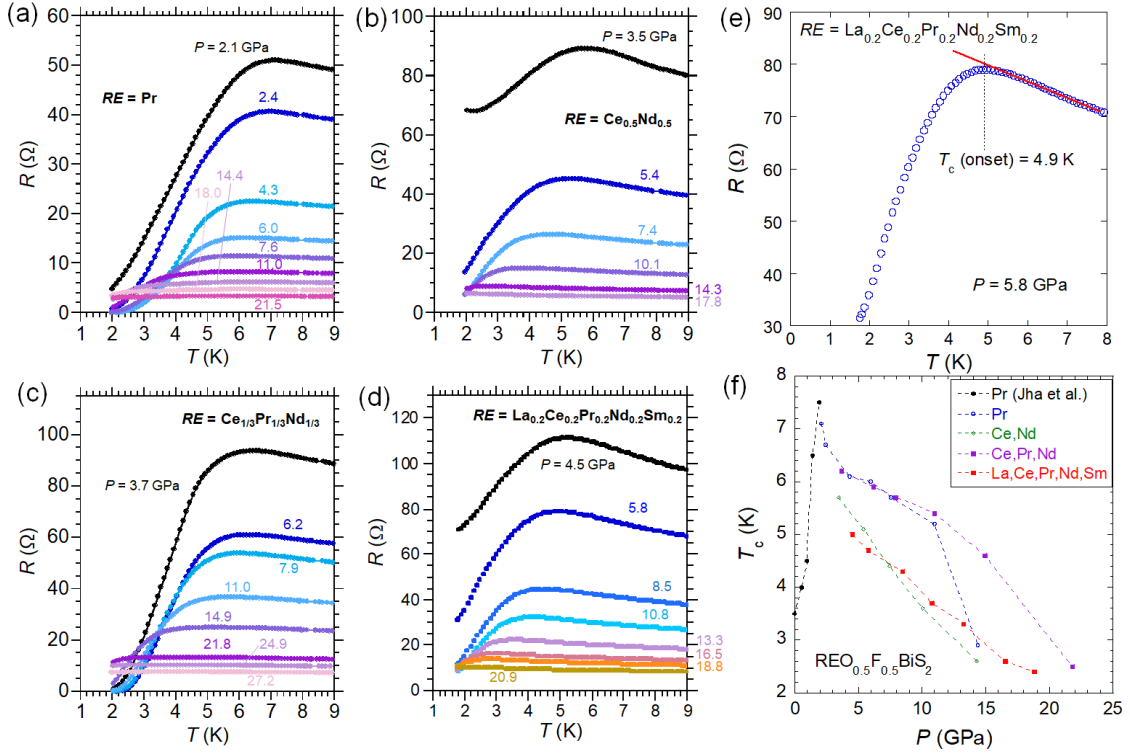

(Color online) Fig. 4. (a–d) Temperature dependences of electrical resistance ($R$) under high pressures for $RE$ = Pr, $Ce_{0.5}Nd_{0.5}$, $Ce_{1/3}Pr_{1/3}Nd_{1/3}$, and $La_{0.2}Ce_{0.2}Pr_{0.2}Nd_{0.2}Sm_{0.2}$. (e) Example of the estimation of $T_c$ (onset) for RE = $La_{0.2}Ce_{0.2}Pr_{0.2}Nd_{0.2}Sm_{0.2}$ under 5.8 GPa. (f) Pressure dependences of $T_c$ for all the samples. For $RE$ = Pr, data published in a previous study on low-pressure range by R. Jha et al. [40] has been added for comparison.

We have investigated the pressure dependence of $T_c$ for $REO_{0.5}F_{0.5}BiS_2$ with difference $RE$-site $\Delta S_{mix}$ to examine whether the $T_c$-pinning phenomena occur in the layered $BiS_2$-based crystal structure by measuring electrical resistance ($R$) under high pressure using DAC. Figures 4(a–d) show the $R$-$T$ plots for $RE$ = Pr, $Ce_{0.5}Nd_{0.5}$, $Ce_{1/3}Pr_{1/3}Nd_{1/3}$, and $La_{0.2}Ce_{0.2}Pr_{0.2}Nd_{0.2}Sm_{0.2}$, respectively. For all samples, $R$ systematically decreases with increasing pressure, which is consistent with pressure studies [37–39]. The residual resistance observed below the superconducting transition is due to connectivity between the sample and DAC electrodes. In such a case, in DAC experiments, we evaluate pressure evolution of superconductivity using onset of the superconducting transition. Here, $T_c$ (onset) was defined as a temperature where the $R$ deviates from a linear $T$ dependence just above the superconducting transition [Fig. 4(e)]. This criterion was utilized in this study because the sharpness of the transitions varied with pressure and compositions. We tested other methods for estimation of $T_c$, such as those based on the anomaly in $T$ derivative plot and crossing points of two linear fitting lines below and above $T_c$, but those attempts failed to compare $T_c$s in the set of $R$-$T$ data.

Figure 4(f) shows the $P$ dependences of $T_c$ for $RE$ = Pr, $Ce_{0.5}Nd_{0.5}$, $Ce_{1/3}Pr_{1/3}Nd_{1/3}$, and $La_{0.2}Ce_{0.2}Pr_{0.2}Nd_{0.2}Sm_{0.2}$. For $RE$ = Pr, low-pressure $T_c$s [40] are plotted together with high-pressure data to overview the $T_c$-$P$ trend. In $REO_{0.5}F_{0.5}BiS_2$, $T_c$ increases with increasing pressure in the tetragonal structure [37,40], and $T_c$ decreases in the monoclinic structure. This trend is observed in all the measured samples with $RE$ = Pr, $Ce_{0.5}Nd_{0.5}$, $Ce_{1/3}Pr_{1/3}Nd_{1/3}$, and $La_{0.2}Ce_{0.2}Pr_{0.2}Nd_{0.2}Sm_{0.2}$. Here, we do not find any effects of the change in $\Delta S_{mix}$ on the $T_c$-$P$ trends, with which we conclude that the $T_c$-pinning phenomenon is absent in HEA-type $REO_{0.5}F_{0.5}BiS_2$.

## 5. Discussion and conclusion

Here, we discuss the reason for the absence of $T_c$-pinning phenomenon in HEA-type $REO_{0.5}F_{0.5}BiS_2$ with a layered crystal structure. According to a previous study on isotope effect on $T_c$ [41], phonon-mediated superconductivity is expected in the monoclinic (high-pressure) phases of $RE$(O,F)$BiS_2$. If modification of phonon characteristics is induced due to the presence of the HEA-type $RE$ site, superconducting properties of $RE$(O,F)$BiS_2$ should be largely affected. In Ref. 25, we reported that the increase in $RE$-site $\Delta S_{mix}$ suppresses in-plane S-site disorder (in-plane displacement parameter $U_{11}$) in $REO_{0.5}F_{0.5}BiS_2$. In such situation, atomic vibration characteristics of the $BiS_2$ conducting layers should not be modified, and glassy atomic vibration, which was observed in HEA-type $M$Te [21], could not be induced in HEA-type $REO_{0.5}F_{0.5}BiS_2$. As shown in Fig. 1, the crystal structures are clearly different between $M$Te (simple three-dimensional NaCl-type structure) and $REO_{0.5}F_{0.5}BiS_2$ (two-dimensional layered structure). Furthermore, the affections of the disorder at the HEA-type site to electronic and vibrational properties would be depending on the position of the HEA-type site. The pressure studies on $M$Te and $REO_{0.5}F_{0.5}BiS_2$ suggests that the incorporation of HEA-type sites is not enough to induce $T_c$-pinning phenomena, and glassy characteristics of atomic vibration would be the direct cause of the $T_c$-pinning phenomena in HEA-type superconductors.

In conclusion, we have studied the crystal structure evolution of HEA-type $REO_{0.5}F_{0.5}BiS_2$ under high pressures and confirmed that the HEA-type sample also exhibits a tetragonal-monoclinic transition at a pressure range similar to that for zero-entropy $LaO_{0.5}F_{0.5}BiS_2$. Therefore, the increase in $RE$-site $\Delta S_{mix}$ does not affect pressure-structure phase diagram in $REO_{0.5}F_{0.5}BiS_2$. To investigate whether $T_c$-pinning phenomena occur in HEA-type $REO_{0.5}F_{0.5}BiS_2$ under high pressures, we measured $P$ dependence of $T_c$ by collecting R-T data for $RE$ = Pr, $Ce_{0.5}Nd_{0.5}$, $Ce_{1/3}Pr_{1/3}Nd_{1/3}$, and $La_{0.2}Ce_{0.2}Pr_{0.2}Nd_{0.2}Sm_{0.2}$. All samples exhibited clear decreases in $T_c$ with increasing pressure in the monoclinic phase, which indicates the absence of $T_c$-pinning phenomena in HEA-type $REO_{0.5}F_{0.5}BiS_2$. The obtained results are consistent with the scenario that $T_c$-pinning phenomena occur when glassy atomic vibrations are induced by introduction of HEA-type sites in HEA-type superconductors.

**Acknowledgment**

The authors thank O. Miura for supports in experiments. The work was partly supported by Tokyo Government Advanced Research (H-31-1) and JSPS-KAKENHI (21H00151).

*E-mail: mizugu@tmu.ac.jp

**References**

1) M. H. Tsai and J. W. Yeh, Mater. Res. Lett. 2, 107 (2014).
2) J. W. Yeh, S. K. Chen, S. J. Lin, J. Y. Gan, T. S. Chin, T. T. Shun, C. H. Tsau, S. Y. Chang, Adv. Energy Mater. 6, 299 (2004).
3) H. Inui, K. Kishida, Z. Chen, Mater. Trans. 63, 394 (2022).
4) P. Koželj, S. Vrtnik, A. Jelen, S. Jazbec, Z. Jagličić, S. Maiti, M. Feuerbacher, W. Steurer, and J. Dolinšek, Phys. Rev. Lett. 113, 107001 (2014).
5) L. Sun, R. J. Cava, Phys. Rev. Mater. 3, 090301 (2019).
6) J. Kitagawa, S. Hamamoto, N. Ishizu, Metals 10, 1078 (2020).
7) S. Marik, M. Varghese, K. P. Sajilesh, D. Singh, and R. P. Singh, J. Alloys Compd. 695, 3530 (2017).
8) K. Stolze, F. A. Cevallos, T. Kong, and R. J. Cava, J. Mater. Chem. C, 6, 10441 (2018).
9) Y. Yuan, Y. Wu, H. Luo, Z. Wang, X, Liang, Z. Yang, H. Wang, X. Liu, and Z. Lu, Front. Mater. 5, 72 (2018).
10) Sourav Marik, Kapil Motla, Maneesha Varghese, K. P. Sajilesh, Deepak Singh, Y. Breard, P. Boullay, R. P. Singh, Phys. Rev. Materials 3, 060602 (2019).
11) N. Ishizu, J. Kitagawa, Results in Phys. 13, 102275 (2019).
12) J. Kitagawa, K. Hoshim Y. Kawasaki, R. Koga, Y. Mizuguchi, T. Nishizaki, J. Alloy Compd. 924, 166473 (2022).
13) J. Guo, H. Wang, F. von Rohr, Z. Wang, S. Cai, Y. Zhou, K. Yang, A. Li, S. Jiang, Q. Wu, R. J. Cava, and L. Sun, Proc. Natl. Acad. Sci. 114, 13144 (2017).
14) C. Huang, J. Guo, J. Zhang, K. Stolze, S. Cai, K. Liu, H. Weng, Z. Lu, Q. Wu, T. Xiang, R. J. Cava, L. Sun, Phys. Rev. Materials 4, 071801 (2020).
15) Y. Mizuguchi, J. Phys. Soc. Jpn. 88, 124708 (2019).
16) Md. R. Kasem, K. Hoshi, R. Jha, M. Katsuno, A. Yamashita, Y. Goto, T. D. Matsuda, Y. Aoki, Y. Mizuguchi, Appl. Phys. Express 13, 033001 (2020).
17) Y. Li, C. Lin, H. Li, X. Li, J. Liu, High Press. Res. 33, 713 (2013).

18) Y. Fujii, K. Kitamura, A. Onodera, Y. Yamada, Solid State Commun. 49, 135 (1984).

19) Y. Bencherif, A. Boukra, A. Zaoui, M. Ferhat, Mater. Chem. Phys. 126, 707 (2011).

20) Md. R. Kasem, Y. Nakahira, H. Yamaoka, R. Matsumoto, A. Yamashita, H. Ishii, N. Hiraoka, Y. Takano, Y. Goto, Y. Mizuguchi, Sci. Rep. 12, 7789 (2022).

21) Y. Mizuguchi, H. Usui, R. Kurita, K. Takae, Md. R. Kasem, R. Matsumoto, K. Yamane, Y. Takano, Y. Nakahira, A. Yamashita, Y. Goto, A. Miura, C. Moriyoshi, Mater. Today Phys. 32, 101019 (2023).

22) Y. Mizuguchi, J. Phys. Soc. Jpn. 88, 041001 (2019).

23) R. Sogabe, Y. Goto, and Y. Mizuguchi, Appl. Phys. Express 11, 053102 (2018).

24) Y. Mizuguchi, K. Hoshi, Y. Goto, A. Miura, K. Tadanaga, C. Moriyoshi, Y. Kuroiwa, J. Phys. Soc. Jpn. 87, 023704 (2018)

25) R. Sogabe, Y. Goto, T. Abe, C. Moriyoshi, Y. Kuroiwa, A. Miura, K. Tadanaga, Y. Mizuguchi, Solid State Commun. 295, 43 (2019).

26) Y. Shukunami, A. Yamashita, Y. Goto, Y. Mizuguchi, Physica C 572, 1353623 (2020).

27) Y. Fujita, K. Kinami, Y. Hanada, Y. Hanada, M. Nagao, A. Miura, S. Hirai, Y. Maruyama, S. Watauchi, Y. Takano, I. Tanaka, ACS Omega 5, 16819 (2020).

28) K. Wang, Q. Hou, A. Pal, H. Wu, J. Si, J. Chen, S. Yu, Y. Chen, W. Lv, J. Y. Ge, S. Cao, J. Zhang, Z. Feng, J. Supercond. Nov. Magn. 34, 1379 (2021).

29) T. Ying, T. Yu, Y. S. Shiah, C. Li, J. Li, Y. Qi, H. Hosono, J. Am. Chem. Soc. 143, 7042 (2021).

30) C. S. Zha, H. K. Mao, R. J. Hemley, Proc. Natl. Acad. Sci. 97, 13494 (2000).

31) Y. Seto, Rev. high press. sci. technol. 20, 269 (2010) (DOI: 10.4131/jshpreview.20.269).

32) F. Izumi, K. Momma, Three-dimensional visualization in powder diffraction, Solid State Phenom. 130, 15 (2007).

33) K. Momma, F. Izumi, J. Appl. Crystallogr. 41, 653 (2008).

34) R. Matsumoto, Y. Sasama, M. Fujioka, T. Irifune, M. Tanaka, T. Yamaguchi, H. Takeya, Y. Takano, Rev. Sci. Instrum. 87, 076103 (2016).

35) R. Matsumoto, A. Yamashita, H. Hara, T. Irifune, S. Adachi, H. Takeya, Y. Takano, Appl. Phys. Express 11, 053101 (2018).

36) G. J. Piermarini, S. Block, J. D. Barnett, R. A. Forman, J. Appl. Phys. 46, 2774 (1975).

37) T. Tomita, M. Ebata, H. Soeda, H. Takahashi, H. Fujihisa, Y. Gotoh, Y. Mizuguchi, H. Izawa, O. Miura, S. Demura, K. Deguchi, Y. Takano, J. Phys. Soc. Jpn. 83, 063704 (2014).

38) Y. Fang, C. T. Wolowiec, D. Yazici, M. B. Maple, Nov. Supercond. Mater. 1, 79 (2015).

39) C. T. Wolowiec, B. D. White, I. Jeon, D. Yazici, K. Huang, M. B. Maple, J. Phys.: Condens. Matter 25, 422201 (2013).

40) R. Jha, H. Kishan, V. P. S. Awana, J. Phys. Chem. Solids 84, 17 (2015).

41) A. Yamashita, H. Usui, K. Hoshi, Y. Goto, K. Kuroki, Y. Mizuguchi, Sci. Rep. 11, 230 (2021).